\documentclass[preprint,10pt]{elsarticle}
\usepackage{amsmath,amssymb}
\usepackage{amscd}
\usepackage{mathrsfs}
\usepackage{float}
\newtheorem{theorem}{Theorem}[section]

\newproof{pf}{Proof}

\numberwithin{equation}{section}
\usepackage{color}
\usepackage{graphicx}
\usepackage{caption}
\usepackage{subcaption}

\def\W{W}

\def\RE{\mathbb R}
\def\CO{{\mathbb C}}

\begin{document}
\begin{frontmatter}
\title{
A quantum hybrid with a thin antenna at the vertex of a wedge}
\cortext[cor]{Corresponding author}
\author[R. Carlone]{Raffaele Carlone\corref{cor}}
\ead{raffaele.carlone@unina.it}
\address[R. Carlone]{Universit\`{a} ``Federico II'' di Napoli, Dipartimento di Matematica e Applicazioni ``R. Caccioppoli'', MSA, via Cinthia, I-80126, Napoli, Italy.}

\author[A. Posilicano]{Andrea Posilicano}
\ead{andrea.posilicano@uninsubria.it}
\address[A. Posilicano]{DiSAT, Universit\`a dell'Insubria, via Valleggio 11, I-22100, Como, Italy}

\begin{abstract}
We study the spectrum, resonances and scattering matrix of a quantum hamiltonian on a ``hybrid surface'' consisting of a half-line attached by its
endpoint to the vertex of a concave planar wedge. At the boundary of the wedge, outside the vertex,  Dirichlet boundary are imposed. The system is tunable by varying the measure of the angle at the vertex.
\end{abstract}
\end{frontmatter}

\section{Introduction}
A \textit{quantum hybrid} is defined as a manifold  obtained by gluing surfaces of different dimensionality. On this manifold the dynamics is described by a Schr\"odinger equation. This kind of models have had considerable success in the literature for many reasons: they are descriptive models for quantum dynamics of microscopic mechanisms as for example semiconductors or quantum dots (see \cite{ETV}), for the analysis of a microwave cavity (see \cite{ES97}), and for  models of quantum point-contact spectroscopy (see \cite{ES86}, \cite{ES87}).
\par
From the theoretical point of view the hybrids are interesting because the conduction of quantum particles inside the hybrid is strictly related to  the formation of quantum resonances, as in the case of quantum wires. Starting from some simplified models (see \cite{ES94})  the development and refinement of spectral analysis techniques has allowed the possibility to consider more and more complicated structure or interactions (see \cite{BGMP02}, \cite{ETV}, \cite{ES87}, \cite{CP11}). In this paper we apply results obtained in the case of models with point interactions and formation of resonances (see \cite{CCF08},\cite{CCF10}, \cite{CCF11}).
\par 
In the model presented in this paper, we  first consider the spectral characterization for a wedge with a convex angle and Dirichlet boundary condition. Using the results in \cite{P13} we  obtain a family of self-adjoint non-Friedrichs Dirichlet Laplacians realizations corresponding to Dirichlet boundary conditions at all the points of the boundary except the corner. This allows to create a quantum hybrid with a half-line attached at the vertex of the not convex corner, using standard technique of the theory of self-adjoint extensions of symmetric operators. The concentration of the non-Friedrichs eigenvectors around the vertex imply the formation of resonances and non  trivial scattering properties.
\par
Let us remark that this phenomenon is almost singular due to the Dirichlet boundary condition for the wedge and  that all the dynamical parameters, as the transmission and reflection coefficient, will depend explicitly on the parameter $1/2\le\beta<1$ related to the measure $\omega>\pi$ of the interior angle of the wedge by $\beta=\pi/\omega$. This makes this device ``tunable'' in the sense that the transitions from the half-line to the plane can be modulated by a variation of $\beta$.

\section{The wedge}
In this section, following \cite[Section 2]{P13} (also see \cite[Section 10.1]{DM}), we recall the definitions of the family of  self-adjoint non-Friedrichs Dirichlet Laplacians  for a non-convex (i.e. $1/2\le \beta<1$) planar wedge 
$$
\W=\{(x_{1},x_{2})\equiv(r\cos\theta,r\sin\theta)\,:\, 0< r<1\,,\ 0<\theta<\pi/\beta\}\,.
$$
Let $\Delta_\W^\circ$ denote the linear given by the restriction of $\Delta_{\W}$, the distributional Laplacian on $\W$,  to the domain $$D(\Delta_\W^\circ)=H^{2}(\W)\cap H^{1}_{0}(\W)=
\left\{\psi\in C(\overline\W):\frac{\partial\psi}{\partial x_i}\,, 
\frac{\partial^{2}\psi}{\partial x_i\partial x_j}\in L^{2}(\W)\,,\  \psi(x)=0,\,  x\in\partial\W\right\}\,.
$$
By Green's formula, $\Delta_\W^\circ$ is clearly symmetric; however, since $W$ is not convex, $\Delta_\W^\circ$ is not self-adjoint (see \cite{BS}).  Indeed the kernel of $(\Delta_\W^\circ)^*$ is one dimensional and is generated by the function 
$$
G(r,\theta)=\frac{1}{\sqrt{\pi}}\,
\left(r^{\beta}-\frac{1}{r^{\beta}}\right)\sin\beta\,\theta\,, 
$$
the unique 
(up to the multiplication by a constant) solution of the boundary value problem
\begin{align*}
\Delta G(x)&=0\,,\quad x\in\W\,,\\
\lim_{y\to x}G(y)&=0\,,\quad x\in\partial\W\backslash\{0\}\,.
\end{align*}
For later convenience we also consider $G_{z}$, the solution of the boundary value problem  
\begin{align*}
(\Delta +z)G_z(x)&=0\,,\quad x\in\W\,,\\
\lim_{y\to x}G_z(y)&=0\,,\quad x\in\partial\W\backslash\{0\}\,.
\end{align*}
For any $z\in\CO$ such that $J_{\beta}(z\,)\not=0$, $J_\nu$ denoting the Bessel function of order $\nu$, $G_{z}$ is given by
\begin{align*}
G_z(r,\theta)
=
\frac{1}{\sqrt{\pi}}\,\Gamma\left(1-\beta\right)\,\left(\frac{\sqrt z}{2}\right)^{\beta}\,
\left(\frac{J_{-\beta}(\sqrt z \,)}{J_\beta(\sqrt z\,)}\,
J_{\beta}(\sqrt z \,r)-J_{-\beta}(\sqrt z \,r)\right)\sin\beta\,\theta\,,
\end{align*}
where $\Gamma$ denotes Euler's gamma function. Here the constants are chosen in order to have $G_z\to G$ as $z\to 0$ and we use the determination of $\sqrt{z}$ such that 
$\text{\rm Im}(\sqrt z)>0$. 
By 
$$
J_\nu(z)=\left(\frac{z}{2}\right)^\nu\,\frac{\tilde J_\nu(z^2)}{\Gamma(1+\nu)}\,,\quad
\tilde J_\nu(z):=1+\sum_{k=1}^\infty\frac{(-1)^k}{k!(k+\nu)!}\,
\left(\frac{z}{4}\right)^k
$$
(here $(k+\nu)!:=(1+\nu)\cdots(k+\nu)$), one has
$$
G_z(r,\theta)=\frac{1}{\sqrt{\pi}}\,
\left(\frac{\tilde J_{-\beta}(z)}{\tilde J_\beta(z)}\,\tilde J_{\beta}(z r^2)\,r^\beta-\tilde J_{-\beta}(z r^2)\,\frac{1}{r^\beta}\right)\sin\beta\,\theta
$$
and so 
\begin{equation}\label{G}
(G-G_{z})(r,\theta)=\frac{1}{\sqrt{\pi}}\,
\left(\left(1-\frac{\tilde J_{-\beta}(z)}{\tilde J_\beta(z)}\,\tilde J_{\beta}(z r^2)\right)\,r^\beta-\left(1-\tilde J_{-\beta}(z r^2)\right)\,\frac{1}{r^\beta}\right)\sin\beta\,\theta\,.
\end{equation}
The set Ext$(-\Delta_\W^\circ)$ of self-adjoint extensions of $-\Delta_\W^\circ$ is given by 
$$
\text{Ext}(-\Delta_\W^\circ)=\{-\Delta^{F}_{\W}\}\cup
\{-\Delta_{\W}^{\alpha}\}_{\alpha\in\RE}\,.
$$
Here $-\Delta^{F}_{W}$ denotes the Friedrichs extension of $-\Delta^{\circ}_{W}$, i.e. $-\Delta^{F}_{W}=(-\Delta_{W})|D(-\Delta_{W}^{F})$, where 
\begin{equation}\label{fe}
D(-\Delta_{W}^{F})=\{\psi_{0}\in C(\overline\W):\|\nabla \psi_{0}\|,\Delta_{W}\psi_{0}\in L^{2}(\W),\, \psi_0(x)=0,\,  x\in\partial\W\}
\end{equation} 
and 
\begin{equation}\label{nfe}
\begin{split}
D(-\Delta_{W}^{\alpha})=&\{\psi\in L^{2}(\W):\psi=\psi_{0}+\xi_{\psi} G\,,\psi_{0}\in D(-\Delta^{F}_{W}),\,
\tau \psi_{0}+(\alpha+1)\xi_{\psi}=0\}\\
=&\{\psi\in L^{2}(\W):\psi=\psi_{\circ}+\xi_{\psi} S\,,\psi_{\circ}\in D(-\Delta^{\circ}_{W}),\,
\tau \psi_{\circ}+\alpha\xi_{\psi}=0\}\,,
\end{split}
\end{equation}
where 
$S(r,\theta):=\frac1{\sqrt\pi}\ r^{\beta}\sin\beta\theta$ 
and $\tau$ denotes the renormalized trace (evaluation) at the vertex
$$\tau\varphi:=\frac{1}2\,\sqrt\pi\beta(\beta+2)\,\lim_{r\downarrow 0}\,
\frac{1}{r^{\beta}}\int_W\varphi(r\textbf{x})\,d\textbf{x}\,.
$$
The operator $-\Delta_{W}^{\alpha}$ acts on its domain as the distributional Laplacian, 
$$
-\Delta_{W}^{\alpha}\psi=-\Delta_{W}\psi\equiv -\Delta^{F}_{W}\psi_{0}\,.
$$
Notice that $D(-\Delta^\alpha_\W)\subset C(\overline W\backslash\{0\})$ and $\psi(x)=0$ whenever $x\in\partial\W\backslash\{0\}$ and $\psi\in D(-\Delta^\alpha_\W)$. 
Using $G$ and $G_{z}$ we define  the Kre\u\i n $Q$-function for the wedge by
$$
Q_z^W:=z\langle G,G_z\rangle_{L^2(\W)}-1\,.
$$
Since $z\langle G,G_z\rangle_{L^2(\W)}=\tau(G-G_{z})$ (see e.g. \cite[equation (A.2)]{P13}), by \eqref{G} one gets
\begin{equation}\label{q}
Q_z^W=-\frac{\tilde J_{-\beta}(z)}{\tilde J_{\beta}(z)}=-\frac{\Gamma(1-\beta)}{\Gamma(1+\beta)}\,\left(\frac{z}{4}\right)^{\beta}\frac{J_{-\beta}(\sqrt z\,)}{J_{\beta}(\sqrt z\,)}
=\frac{\Gamma(-\beta)}{\Gamma(\beta)}\,\left(\frac{z}{4}\right)^{\beta}\frac{J_{-\beta}(\sqrt z\,)}{J_{\beta}(\sqrt z\,)}\,.
\end{equation} 
For any $z\in \rho(-\Delta^{\alpha}_{W})\cap \rho(-\Delta^{F}_{W})$ the kernel  of the resolvent $(\Delta_{W}^{\alpha}+z)^{-1}$ is given by 
\begin{equation}\label{resolvent}
R_{z}^\alpha(r,\theta;r',\theta')=R_z^F(r,\theta;r',\theta')
-\left(\alpha-Q_z^{W}\right)^{-1} 
G_z(r,\theta)G_z(r',\theta')\,,
\end{equation}
where $R^F_z$ denotes the resolvent kernel of  the Friedrichs-Dirichlet Laplacian $-\Delta^F_\W$, i.e. 
$$
R^F_z(r,\theta;r',\theta'):=\sum_{m,n\ge 1}\frac{\psi_{m,n}(r,\theta)\psi_{m,n}(r',\theta')}{-\lambda_{m,n\beta}^2+z}\,.
$$ 
Here $\lambda_{m,\nu}$ denotes the $m$-th strictly positive zero of $J_{\nu}$ and 
$$
\psi_{m,n}(r,\theta)=2\,\sqrt{\frac{\beta}{\pi}}\,\frac{J_{n\beta}(\lambda_{m,n\beta}r)}{J_{n\beta+1}(\lambda_{m,n\beta})}\,\sin n\beta\theta
$$
is the normalized eigenfunctions of  $-\Delta^F_\W$ corresponding to the eigenvalue $\lambda^{2}_{m,n\beta}$. Notice that,  since $J_{\beta}$ has only real zeroes, both $G_{\lambda}$ and $Q^W_{\lambda}$ are well defined for any $\lambda\not=\lambda^{2}_{m,\beta}$, $m\ge 1$. 
\par 
Since $\tau\psi_{m,n}\not=0$ if and only if $n=1$, by \cite[Example 3.1]{P2014}, one gets
$$
\Sigma_{F}:=\sigma(-\Delta_{W}^{\alpha})\cap \sigma(-\Delta_{W}^{F})=\{\lambda_{m,n\beta}^2\,,\ m\ge 1\,,n>1\}\,.
$$
Thus
\begin{equation}\label{eiwedge}
\sigma(-\Delta_\W^\alpha)=\sigma_{d}(-\Delta_\W^\alpha)
=\Sigma_{\alpha}\cup\Sigma_F\,,
\end{equation}
where
$$
\Sigma_\alpha=\{\lambda: Q^{W}_{\lambda}=\alpha\}=\Sigma^{-}_{\alpha}\cup\Sigma^{+}_{\alpha}
$$
$$
\Sigma^{-}_{\alpha}:=\{\lambda<0: Q^{W}_{\lambda}=\alpha\}=\left\{\lambda<0: \frac{\Gamma(-\beta)}{\Gamma(\beta)}\,\left(\frac{|\lambda|}{4}\right)^{\beta}\frac{I_{-\beta}(\sqrt{ |\lambda|}\,)}{I_{\beta}(\sqrt{|\lambda|}\,)}=\alpha\right\}
$$
(here $I_{\nu}$ denotes the modified Bessel function of the first kind)\,,
$$
\Sigma^{+}_{\alpha}:=\{\lambda\ge 0: Q^{W}_{\lambda}=\alpha\}=\left\{\lambda\ge 0:
\frac{\Gamma(-\beta)}{\Gamma(\beta)}\,\left(\frac{\lambda}{4}\right)^{\beta}\frac{J_{-\beta}(\sqrt \lambda\,)}{J_{\beta}(\sqrt \lambda\,)}=\alpha\right\}\,.
$$
By \cite[relations (A.1) and (A.3)]{P13} one has 
\begin{align*}
z\langle G,G_z\rangle_2-w\langle G,G_w\rangle_2=
\tau(G_w-G_z)
=(z-w)\langle G_w,G_z\rangle_2\,.
\end{align*}
Here and below $\langle\cdot,\cdot\rangle_{2}$ and $\|\cdot\|_{2}$ denote the scalar product and the corresponding norm in $L^{2}(W)$. Therefore 
\begin{equation}\label{dz}
\frac{d\ }{dz}\, Q^{W}_{z}=\frac{d\ }{dz}\,z\langle G,G_{z}\rangle_2=\|G_{z}\|_2^{2}>0
\end{equation}
and so the map $\lambda\mapsto Q_{\lambda}^{W}$ is strictly increasing on $\RE\backslash\{\lambda^{2}_{m,\beta}\}_{m=1}^{+\infty}$. 
 Since $Q_{0}^{W}=-1$, 
$$\lim_{\lambda\to-\infty}Q_{\lambda}^{W}=-\infty\,,\qquad 
\lim_{\lambda\to\lambda^{2}_{m,\beta}\mp}Q_{\lambda}^{W}=\pm\infty\,,
$$
one has 
$$
\Sigma^{+}_{\alpha}=\Sigma_{\alpha}^{\circ}\cup\{ \lambda_m\}_{m=1}^{+\infty}\,,
\qquad  \lambda_m\in (\lambda^{2}_{m,\beta},\lambda^{2}_{m+1,\beta})\,,
$$
the set $\Sigma^{-}_{\alpha}\cup\Sigma^{\circ}_{\alpha}\subset (-\infty,\lambda^{2}_{1,\beta})$ is a nonempty singleton, $0\in \Sigma^{+}_{\alpha}$ if and only if $\alpha=-1$ and $\Sigma^{-}_{\alpha}$ is nonempty if and only if $\alpha<-1$. Special cases are the following:
$$
\beta=\frac12\quad\Longrightarrow\quad\Sigma^{-}_{\alpha}=\{\lambda<0:-\sqrt\lambda\coth\sqrt\lambda=\alpha\}\,,\quad \Sigma^{+}_\alpha=\{\lambda\ge 0: -\sqrt\lambda\cot\sqrt\lambda=\alpha\}\,;
$$
$$
\alpha=0\quad\Longrightarrow\quad\Sigma^{+}_{0}=\{\lambda^2_{k,-\beta}\}_{k=1}^{+\infty}\,;
$$
$$
\alpha=0\,,\ \beta=\frac12\quad\Longrightarrow\quad\Sigma_0^{+}=\left\{\pi^2\left(k-\frac12\right)^2\right\}_{k=1}^{+\infty}\,.
$$
\section{The Hybrid}
In this section we  analyze the spectral and scattering properties related to a manifold obtained by gluing the wedge with an half-line. This procedure  has  been widely studied in literature; the procedure allowing to paste varieties with different dimensionality  is discussed in several papers (see \cite{ES94,ES87,ES97,ES07,CP11}). 
\par
The peculiarity of the present  model  lays in the fact that the sticking point is the vertex of the wedge, and along its boundary, outside the vertex itself, Dirichlet boundary condition hold. In general, in this configuration,  it is usually impossible to obtain a hybrid with non-trivial properties. However the appearing of resonances  shows that this procedure can be applied. This model shows also an interesting tunability: the opening angle of the wedge, in a suitable range of values, will influence the spectral properties.
\par
The Hilbert space for the hybrid is 
\begin{equation}
\mathcal{H}=L^{2}(0,+\infty)\oplus L^{2}(W)
\end{equation}
The functions in $\mathcal{H}$ will be written as $\Psi=\phi\oplus\psi$. As Hamiltonian on the half-line (the lead) we will take the operator $H^{L}=-L_{N}$, where $L_{N}$ denotes the Laplacian $L\phi:=\phi''$ with Neumann boundary condition at the endpoint.  For the wedge we will consider the Hamiltonian $H^{W}=-\Delta_\W^F$. The kernel of the resolvent $(L_{N}+z)^{-1}$, $z\in \CO\backslash[0,+\infty)$, is given by 
$$
R^{N}_{z}(x,y)=-\frac{i}2\,\frac{e^{i\sqrt{z}\,|x-y|}+e^{i\sqrt{z}\,(x+y)}}{\sqrt{z}}\,,
$$
As before we use the determination of $\sqrt{z}$ such that 
$\text{\rm Im}(\sqrt z)>0$. The corresponding Kre\u\i n $Q$-function for the lead is $$Q^{L}_{z}=\frac{i}{\sqrt{z}}\,.$$ 
For the construction of the hybrid we will follow a standard procedure for the construction of self-adjoint extensions (see e.g. \cite{AGHH}, \cite{P08} for the general theory and \cite{ES86}, \cite{CP11} for hybrids)  starting from the Green function of the decoupled system. We define the boundary maps $\Gamma_k:H^{2}(0,+\infty)\oplus D((\Delta_{W}^{\circ})^{*})\to \CO\oplus \CO$, $k=1,2$, by 
\begin{equation}
\Gamma_{1}(\phi\oplus\psi)=\phi(0_+)\oplus\tau\psi_{0}\,,\quad 
\Gamma_{2}(\phi\oplus\psi)=(-\phi'(0_+))\oplus\xi_{\psi}\,,
\end{equation}
where
$$
D((\Delta_{W}^{\circ})^{*})=\{\psi\in L^{2}(\W):\psi=\psi_{0}+\xi_{\psi} G,\ \psi_{0}\in D(\Delta^{F}_{W}),\ \xi_{\psi}\in\CO\}\,.
$$
Notice that $(\Gamma_{1},\Gamma_{2},\CO^{2})$ is a boundary triple (see e.g. \cite[Section 14]{S}) for the symmetric operator $L^{\circ}\oplus \Delta_{W}^{\circ}$, where $L^{\circ}$ is the minimal realization of $L$, i.e. 
\begin{align*}
D(L^{\circ})=H_{0}^{2}(0,+\infty)
=\{\phi\in L^{2}(0,+\infty)\cap C^{1}[0,+\infty): \phi',\,\phi''\in L^{2}(0,+\infty),\,\,\phi(0_{+})=\phi'(0_{+})=0\}\,.
\end{align*}
Introducing the symmetric matrix 
\begin{equation}\label{teta}
\Theta=\left(\begin{matrix}\gamma & \epsilon \\ \epsilon &\alpha+1\end{matrix}\right)\,,
\end{equation}
and 
the  Kre\u\i n $Q$-function for the hybrid 
\begin{equation}
Q^H_{z}=\left(\begin{matrix}Q^{L}_{z} & 0\\0 & Q^{W}_{z}+1\end{matrix}\right)\,,
\end{equation}
one then obtains the self-adjoint Hamiltonian $H_{\Theta}$ on the hybrid defined by 
$$
D(H_{\Theta}):=\{\Psi=\phi\oplus\psi\in H^{2}(0,+\infty)\oplus D((\Delta_{W}^{\circ})^{*}): 
(\Gamma_{1}+\Theta\,\Gamma_{2})\Psi=0\}\,,\quad 
H_{\Theta}\Psi:=(-\phi'')\oplus(-\Delta_{W}\psi)\,.
$$
and the kernel of the resolvent  $(-H_{\Theta}+z)^{-1}$, $z\in\rho(H_{\Theta})\cap \rho(H^{L}\oplus H^{W})$, is  
\begin{equation}\label{greenhyb}
R^{\Theta}_{z}(x,y;\textbf{x},\textbf{y})=\left(
\begin{matrix}R^{N}_{z}(x,y)& 0\\0 &R_{z}^{F}(\textbf{x},\textbf{y})\end{matrix}\right)
-\left(\begin{matrix}R^{N}_{z}(x,0)& 0\\0 &G_{z}(\textbf{x})\end{matrix}\right)\!
\left(\begin{matrix}\gamma-Q^{L}_{z} & \epsilon\\\epsilon &\alpha-Q^{W}_{z}\end{matrix}\right)
^{\!-1}\!\left(\begin{matrix}R^{N}_{z}(0,y)& 0\\0 &G_{z}(\textbf{y})\end{matrix}\right)\,.
\end{equation}
Let us consider preliminarly the case $\epsilon=0$. From (\ref{greenhyb}) we get the operator $(-L_{\gamma})\oplus(-\Delta_{W}^{\alpha})$ for the decoupled case, where $-L_{\gamma}$ is the self-adjoint operator  describing a point interaction of strength $a=-1/\gamma$ placed at the endpoint of the half-line; the operator is diagonal and there is no interaction between the two components. In this case 
$$\sigma_{ess} ((-L_{\gamma})\oplus(-\Delta^{\alpha}_{W}))=\sigma_{ac} ((-L_{\gamma})\oplus(-\Delta^{\alpha}_{W}))=[0,+\infty)
$$ 
$$\sigma_{d} ((-L_{\gamma})\oplus(-\Delta^{\alpha}_{W}))=\sigma_{d} (-L_{\gamma})\cup \Sigma^{-}_{\alpha}\,,$$ 
$$\sigma_{p} ((-L_{\gamma})\oplus(-\Delta^{\alpha}_{W}))=\sigma_{d} (-L_{\gamma})\cup \sigma(-\Delta^{\alpha}_{W})\,.$$ 
Notice that $\sigma_{d} (-L_{\gamma})=\{-\gamma^{-2}\}$ whenever $a<0$, otherwise $\sigma_{d} (-L_{\gamma})=\emptyset$.
\par
Let us now discuss what happens to the  eigenvalues embedded in the continuos part of the spectrum  after the hybridization. The positive eigenvalues in $\Sigma_{F}$ relative to the Friedrichs extension are unaffected by the hybridization except for the fact that they are embedded in the continuos spectrum of $-L_{N}$. The positive eigenvalues in $\Sigma^{+}_{\alpha}$ are also embedded in the continuous  spectrum and for $\epsilon\neq 0$ they can eventually move from the real line. They will correspond to the solutions of
\begin{equation}\label{nf}
\det \left(\Theta-Q_z^{H}\right)=0
\end{equation}
We resume the spectral properties for the hybrid in the following 
\begin{theorem} The spectrum of $H_{\Theta}$ is given by:
\begin{itemize}
\item[i)] the essential and absolutely continuous spectra $\sigma_{ess}(H_{\Theta})=\sigma_{ac}(H_{\Theta})=[0,\infty)$;
\item[ii)] the discrete spectrum $\sigma_{d}(H_{\Theta})=\{\lambda< 0:(\gamma-Q^{L}_{\lambda})(\alpha-Q^{W}_{\lambda})=\epsilon^{2}\}$;
\item[iii)] the point spectrum $\sigma_{p}(H_{\Theta})=\sigma_{d}(H_{\Theta})\cup\Sigma_{F}$;
\item[iv)] the continuous spectrum $\sigma_{c}(H_{\Theta})=[0,\infty)\backslash\Sigma_{F}$.
\end{itemize}
Moreover 
\begin{itemize}\item[v)] there are infinitely many resonances 
\begin{equation}\label{res}
r_{m}(\epsilon)=\lambda_{m}+
w_{m}\epsilon^{2}+\mathcal{O}( \epsilon ^{3})\,,
\end{equation}
where $\lambda_{m}\in\Sigma^{+}_{\alpha}$ is a  positive non-Friedrichs eigenvalue of $-\Delta^{\alpha}_{W}$ and $w_{m}\in\CO_{-}$ is explicitly given below in \eqref{weakres}. 
\end{itemize}
\end{theorem}
\begin{pf} Since $H_{\Theta}$ is a rank two resolvent perturbation of $(-L_{N})\oplus(-\Delta^{F}_{W})$, the preservation of the essential and absolutely continuous spectrum spectrum is consequence of the Weyl and of Kato-Birman criteria:
\begin{equation*}
\sigma_{ess} (H_{\Theta})=\sigma_{ess} ((-L_{N})\oplus(-\Delta^{F}_{W}))=\sigma_{ess} (-L_{N})=[0,+\infty)\,,
\end{equation*}
\begin{equation*}
\sigma_{ac} (H_{\Theta})=\sigma_{ac} ((-L_{N})\oplus(-\Delta^{F}_{W}))=\sigma_{ac} (-L_{N})=[0,+\infty)\,.
\end{equation*}
From the structure of the resolvent it is clear that the eigenvalues of $-\Delta^{F}_{W}$ in $\Sigma_{F}$ are unperturbed by the formation of the hybrid while the other ones, belonging to $\sigma(-\Delta^{F}_{W})\backslash\Sigma_{F}=\{\lambda^{2}_{m,\beta}\}_{m=1}^{+\infty}$, are in $\rho(H_{\Theta})$ (see \cite[Example 3.1]{P2014}).  The only others points in the  spectrum are given by the singularities of $(\Theta-Q^{H}_{z})^{-1}$, i.e. by the solutions of the equation (\ref{nf}) 
\begin{equation*}
\left(\gamma-Q^{L}_{z}\right) \left(\alpha-Q^{W}_{z}\right)= \epsilon ^{2}\,.
\end{equation*}
By a direct computation one can easily check that there are no square integrable solutions of the equation $H_{\Theta}\Psi=0$, so that $0\notin\sigma_{p}(H_{\Theta})$.
\par 
The wedge eigenvalues of the unperturbed Hamiltonian,
corresponding to $ \epsilon =0$,  are all simple, and since the perturbation is of finite rank in the resolvent sense, continuity of the
singularities with respect to $ \epsilon $ is guaranteed.
\par
Preliminarily we examine qualitatively the solutions of \eqref{nf}. We can rewrite (\ref{nf}) with  $z=\lambda>0$ as
\begin{equation}\label{qual}
\sqrt{\lambda}={i}\ \frac{\alpha -Q^{W}_{\lambda}}
{\gamma\left(\alpha -Q^{W}_{\lambda}\right)- \epsilon ^{2}}
\end{equation}
This equation can not be solved by any real
positive $\lambda$ because on the left-hand side we have always a real positive number while  on the right-hand side is always purely imaginary: any solution of this equation lies  outside the real axis. The singularities of the
resolvent, in the weak coupling regime, can be found as a  series expansion in $ \epsilon $ using a recursive procedure. To begin with we rewrite the equation \eqref{nf} as
\begin{equation}
  Q_{z}^{W}=f(z):=\alpha-\frac{\epsilon^{2}\sqrt z}{\gamma\sqrt z-{i}}
\label{geneve}
\end{equation}
In the next recursive procedure we denote by $\lambda_{m}$ the eigenvalues in $\Sigma^{+}_{\alpha}$ (so that $Q^{W}_{\lambda_{m}}=\alpha$) and we look for the fixed points of the equations (here we use \eqref{dz})
\begin{equation}\label{rec}
{Q^{W}_{z_k}}\simeq
\rho(\lambda_{m})(z_{k}-z_{0})=f(z_{k-1})\,,\qquad 
\rho(\lambda_{m})
=\left.\frac{d}{dz}\right|_{z=\lambda_{m}}{Q^{W}_{z}}=
\|G_{\lambda_{m}}\|^{2}_2>0\,.
\end{equation}
In this way we can determine the position of the singularities in the leading order in $ \epsilon $. We will start from $z_{0}=\lambda_{m}$ and evaluate the  fixed points with the recursive
procedure, from (\ref{rec}).  At the end we  have a positive real part and a negative imaginary part; in a leading order in $ \epsilon $ the resonances $r_{m}(\epsilon)$ are given by \eqref{res}, where 
\begin{equation}
\label{weakres}
\text{\rm Re}(w_{m})=-\frac{\lambda_{m}\gamma}{\|G_{\lambda_{m}}\|^{2}_2(\lambda_{m}\gamma^{2}+1)}  \qquad\text{\rm Im} (w_{m})=-\frac{\sqrt{\lambda_{m}}}{\|G_{\lambda_{m}}\|^{2}_2(\lambda_{m}\gamma^{2}+1)}\,.
\end{equation}
The convergence of the sequence in a ball of radius $ \epsilon ^2$
centered at $z_{0}=\lambda_{m}$ can be proved using the following estimates,
$$
\|z_{1}-z_{0}\|=\frac{1}{\|G_{\lambda_{m}}\|^{2}_2}\ \|f(z_{0})\|<
C_1  \epsilon ^2\,,\qquad
\|z_{k+1}-z_{k}\|\leqslant C_2\,  \epsilon ^2
\|z_{k}-z_{k-1}\|\,,
$$
where the constants $C_1$ and $C_{2}$ are independent of $ \epsilon $.
\end{pf}
The results of the previous theorem are shown in the following  plots obtained numerically.

\begin{figure}[!htb]
\minipage{0.32\textwidth}
  \includegraphics[width=\linewidth]{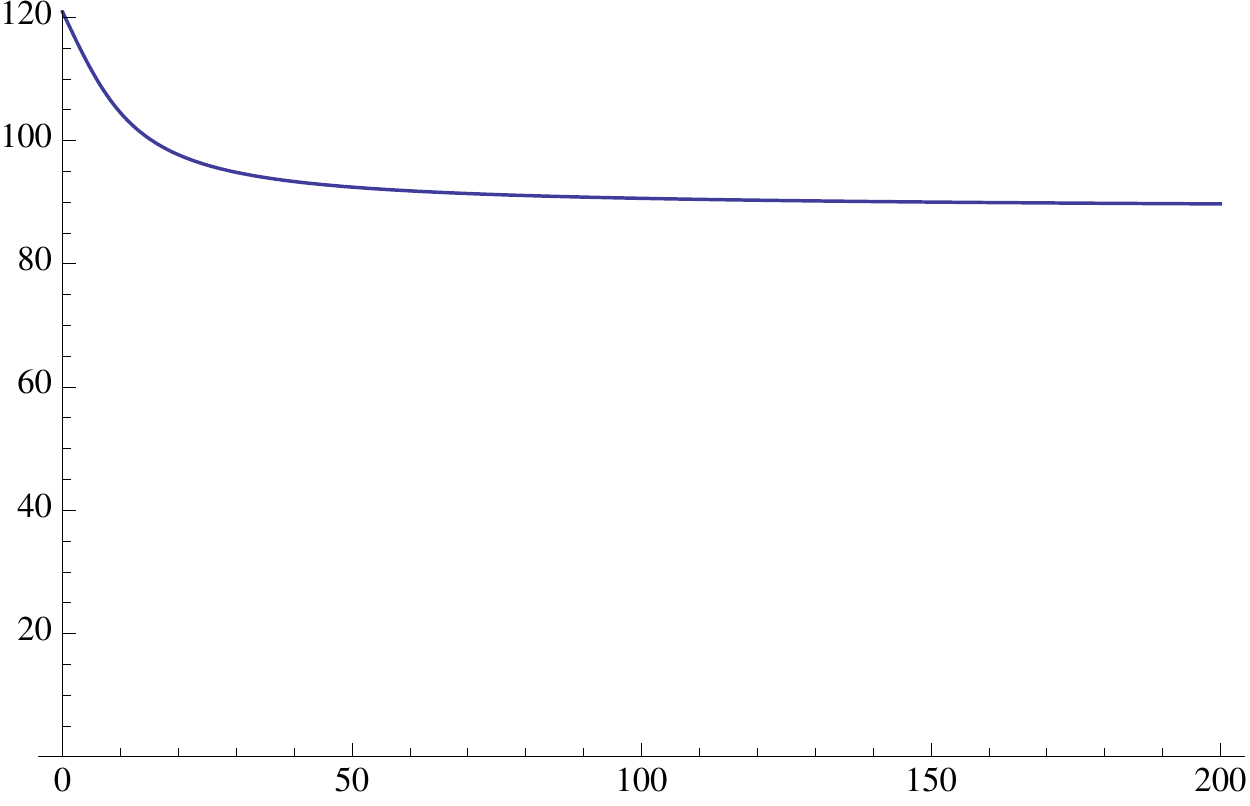}
  \caption{{\footnotesize Real part of a resonance as a function of $ \epsilon $ 
  }}
\label{realpart}
\endminipage\hfill
\minipage{0.32\textwidth}
  \includegraphics[width=\linewidth]{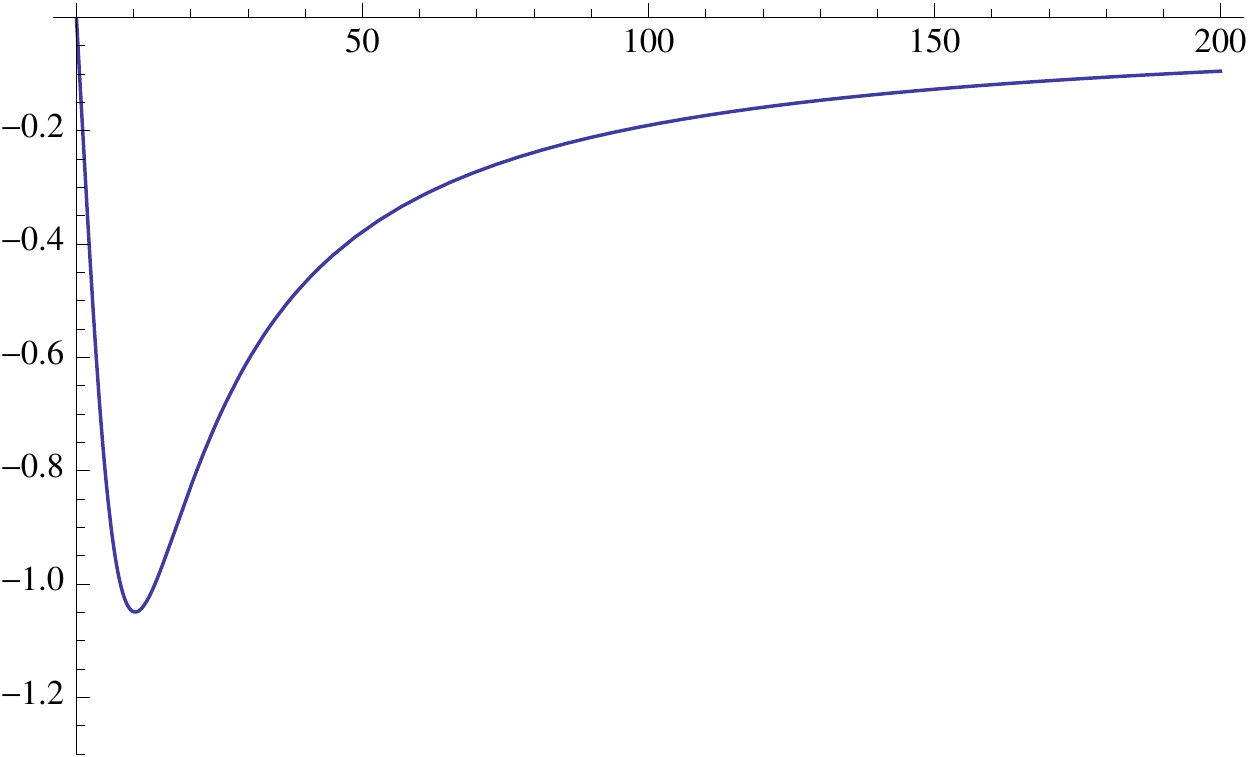}
  \caption{{\footnotesize Imaginary part of a resonance as a function of $ \epsilon $ 
  }}
\label{imlpart}
\endminipage\hfill
\minipage{0.32\textwidth}%
\includegraphics[width=6cm]{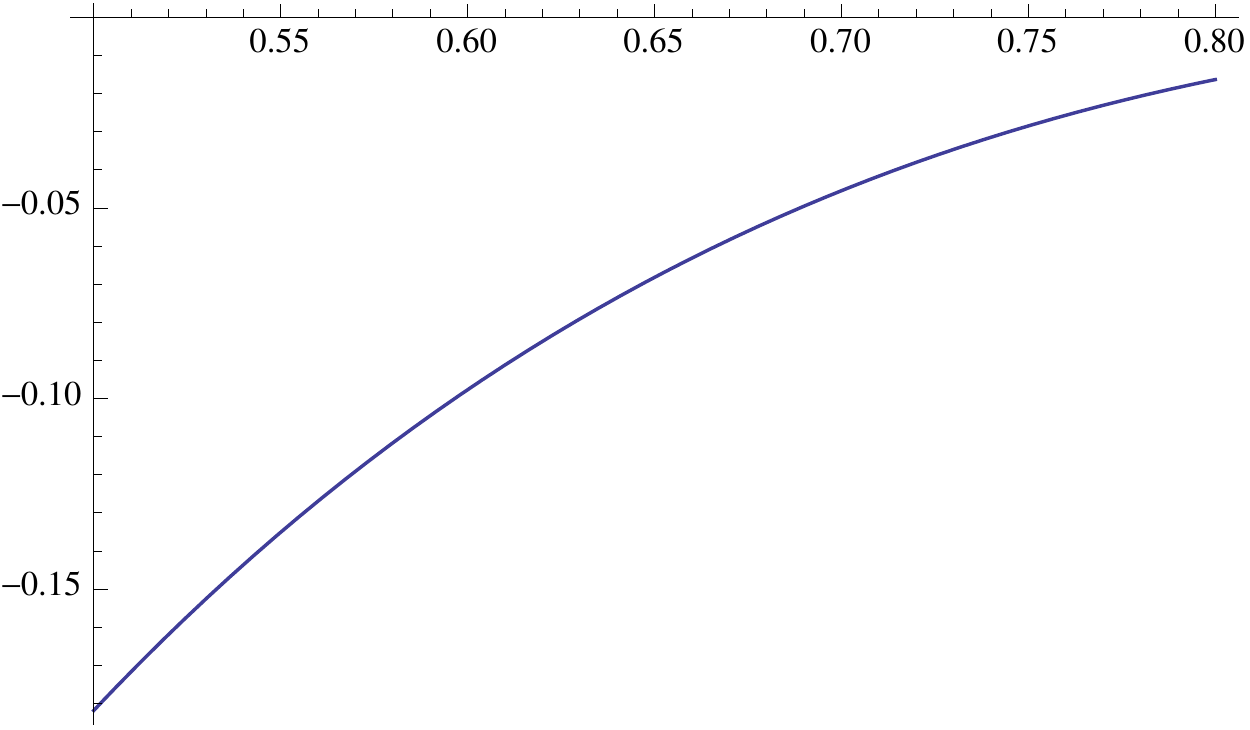}
  \caption{Imaginary part of a resonance as a function of $\beta$ with $ \epsilon =1$}
\label{angledep}\endminipage
\end{figure}

As it is usual in many models involving resonance problems (see e.g.
 \cite{CP11}, \cite{ES94}) for large values of the coupling constant $ \epsilon $ the resonance can approach an eigenvalue again: this is due to the fact that  that $H_{\Theta}$ reduces in the limit
$ \epsilon \rightarrow\infty$ to $(-L_{N})\oplus (-\Delta^{F}_{W})$.
\par
An interesting feature of this model is that,  as evidenced by the (\ref{weakres}) and by  figure (\ref{angledep}), the spectral properties are related to the value of the geometric parameter $\beta$. In this way it is possible to tune the conduction along the hybrid by the variation of $\beta$.
\par
Using the formula for the rank two resolvent difference $(H_{\Theta}+z)^{-1}-((-L_{N})\oplus(- \Delta^{F}_{W})+z)^{-1}$ given in \eqref{greenhyb} and results provided in \cite[Section 4.3.2]{AK} (see also \cite{BMN}), it is possible to write the scattering matrix $\mathcal{S}(\lambda)$, $\lambda>0$, explicitly in terms of the Kre\u\i n $Q$-function and the parameters $\alpha$, $\gamma$, $\beta$ and $\epsilon$:  
\[ \label{scattering_matrix}
\mathcal{S}(\lambda)= \left(\begin{array}{cc}\frac{\left(\gamma+\frac{{i}}{\sqrt{\lambda}}\right)\left(\alpha-Q_{\lambda}^{W}\right)-\epsilon^{2}}{\left(\gamma-\frac{{i}}{\sqrt{\lambda}}\right)\left(\alpha-Q_{\lambda}^{W}\right)-\epsilon^{2}} & 0 \\0 & 1\end{array}\right)=
\left(\begin{array}{cc}\frac{\left(\gamma+\frac{{i}}{\sqrt{\lambda}}\right)\left(\alpha-\frac{\Gamma(-\beta)}{\Gamma(\beta)}\,\left(\frac\lambda{4}\right)^{\beta}\,\frac{J_{-\beta}(\sqrt\lambda)}{J_{\beta}(\sqrt\lambda)}\right)-\epsilon^{2}}{\left(\gamma-\frac{{i}}{\sqrt{\lambda}}\right)\left(\alpha-\frac{\Gamma(-\beta)}{\Gamma(\beta)}\,\left(\frac\lambda{4}\right)^{\beta}\,\frac{J_{-\beta}(\sqrt\lambda)}{J_{\beta}(\sqrt\lambda)}\right)-\epsilon^{2}} & 0 \\0 & 1\end{array}\right)
\,.
\]
Putting $\lambda=k^{2}$, $k>0$, we can describe the reflection amplitude of a particle travelling along the halfline as 

\begin{equation}
 \label{refcoef}
\mathcal{R}(k)=\frac{\left(\gamma+\frac{{i}}{k}\right)\left(\alpha-\frac{\Gamma(-\beta)}{\Gamma(\beta)}\,\left(\frac{k}{2}\right)^{2\beta}\,\frac{J_{-\beta}(k)}{J_{\beta}(k)}\right)-\epsilon^{2}}{\left(\gamma-\frac{{i}}{k}\right)\left(\alpha-\frac{\Gamma(-\beta)}{\Gamma(\beta)}\,\left(\frac{k}{2}\right)^{2\beta}\,\frac{J_{-\beta}(k)}{J_{\beta}(k)}\right)-\epsilon^{2}}\,.
 \end{equation}

\section{Conclusions}
In this paper we have shown that it is possible to build a quantum hybrid consisting of a particular structure that allows to glue an half-line to the edge of a domain with Dirichlet boundary conditions. The results of  this simplified model can be easily generalized to more complicated structures as long as a non convex polygon is present.
\par
From the application point of view   it is interesting the persistence of a geometrical dependence in the scattering matrix and in the reflection amplitude.
\par
Using well established results present in the literature,  it is possible to generalize  the result to the case of finitely many or also infinite number of connected hybrids (see e.g. \cite{ETV}).

\vskip10pt\noindent
{\bf Acknowledgments.} The authors acknowledge the support of the FIR 2013 project ``Condensed Matter in Mathematical Physics'', Ministry of University and
Research of Italian Republic  (code RBFR13WAET).

\end{document}